\documentclass[%
 reprint,pre,
showpacs,
 amsmath,amssymb,superscriptaddress,
 aps
]{revtex4-1}

\usepackage{amsmath}
\usepackage{amssymb}
\usepackage{graphicx}
\usepackage{dcolumn}
\usepackage{bm}
\usepackage{hyperref}
\usepackage{verbatim}
\usepackage{color}
\usepackage{makebox}
\usepackage{pdfpages}

\usepackage{verbatim}
\usepackage{graphics}
\usepackage{longtable} 
\usepackage{subfigure}
\usepackage{amsmath}
\usepackage{wrapfig}
\usepackage{epsfig}
\usepackage[miktex]{gnuplottex}
\usepackage{float}
\usepackage{graphicx}
\usepackage{array}
\usepackage{psfrag}
\usepackage{color}
\usepackage{bbold}

\usepackage{multirow}

\usepackage{tikz}

\renewcommand{\vec}[1]{\mathbf{#1}}

\newcommand\Figure[5][\fullwidth]{%
  \label{\thefigure:WideFigure}
  \begin{figure}[!htb]%
    \abovecaptionskip=0pt\belowcaptionskip=0pt
    \ifthenelse{\isodd{\pageref{\thefigure:WideFigure}}}{}{\hspace*{-\marginparwidth}}%
    \begin{minipage}{#1}
     \begin{minipage}[b]{0.45\linewidth}\centering
     \includegraphics[width=0.95\linewidth]{#2}
     \caption{#3}
     \end{minipage}\hfill%
     \begin{minipage}[b]{0.45\linewidth}\centering
     \includegraphics[width=0.95\linewidth]{#4}
     \caption{#5}
     \end{minipage}%
  \end{minipage}%
 \end{figure}}

\begin{document}

\title{Formation and Dissolution of Bacterial Colonies}

\author{Christoph A. Weber} 
\affiliation{Max Planck Institute for the Physics of Complex Systems, N\"othnitzer Str.\ 38, 01187 Dresden, Germany}
\author{Yen Ting Lin} 
\affiliation{Max Planck Institute for the Physics of Complex Systems, N\"othnitzer Str.\ 38, 01187 Dresden, Germany}
\author{Nicolas Biais}
\affiliation{Department of Biology, Brooklyn College, City University of New York, Brooklyn, NY 11210}
\author{Vasily Zaburdaev}
\affiliation{Max Planck Institute for the Physics of Complex Systems, N\"othnitzer Str.\ 38, 01187 Dresden, Germany}


\begin{abstract}
Many organisms form colonies for a transient period of time to withstand environmental pressure. Bacterial biofilms are a prototypical example of such behavior. Despite significant interest across disciplines, physical mechanisms governing the formation and dissolution of bacterial colonies are still poorly understood. Starting from a kinetic description of motile and interacting cells we derive a hydrodynamic equation for their density on a surface. We use it to describe formation of multiple colonies 
with sizes consistent with experimental data and to discuss their dissolution.
\end{abstract}

\maketitle



\begin{figure}[bt]
\centering
\begin{tabular}{cc}
\includegraphics[width=0.25\textwidth]{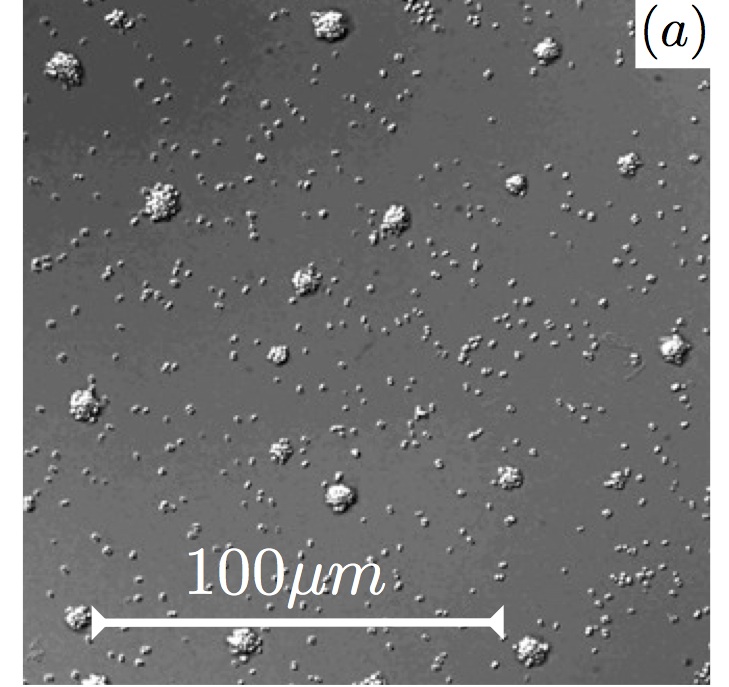} &
\includegraphics[width=0.25\textwidth]{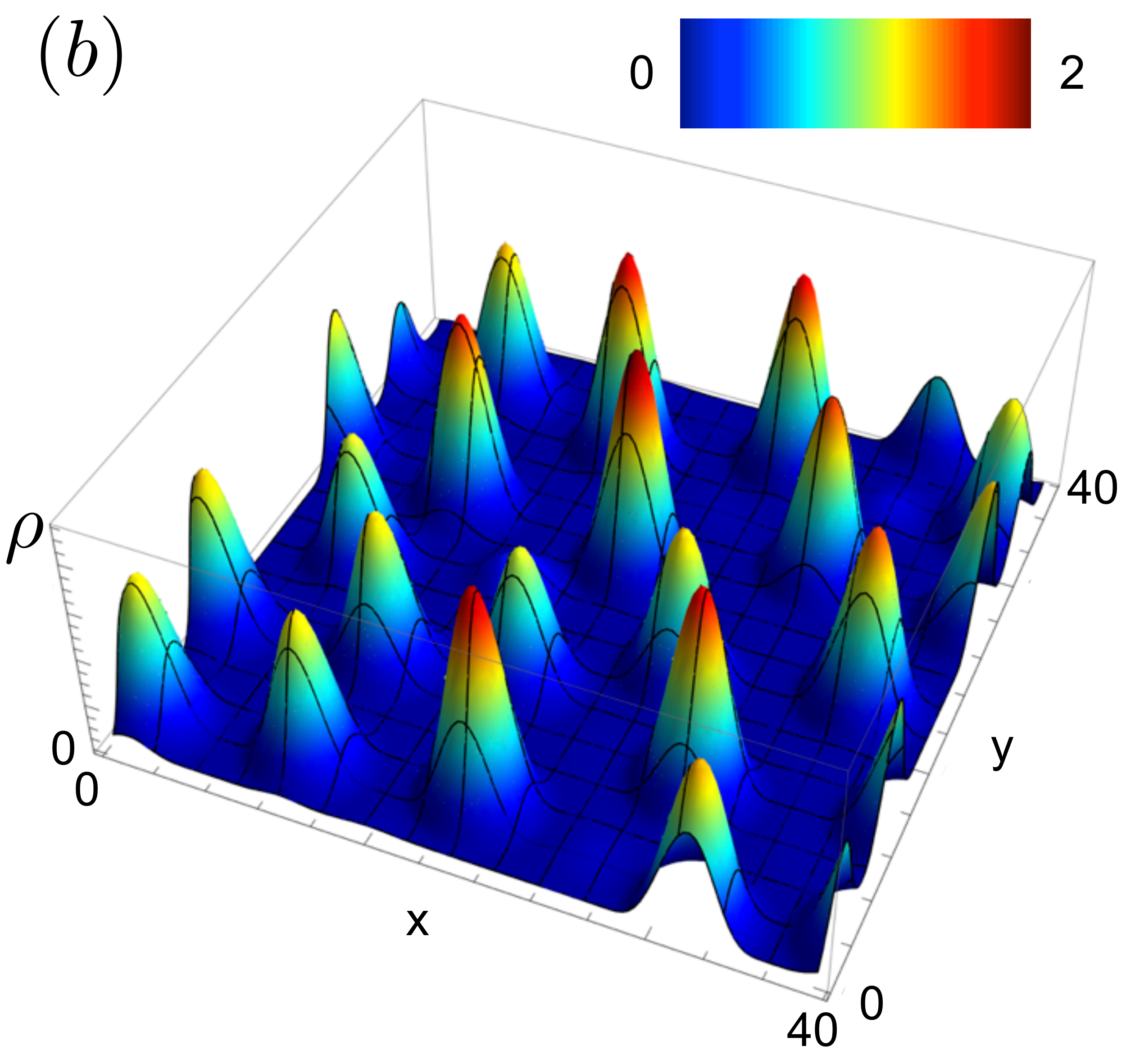}\\
\end{tabular}
\caption{\label{fig:snapshots}  (color online)
\text{(a)} \emph{N. gonorrhoeae}  colonies  three hours after sedimentation on a plastic substrate.
\text{(b)} Typical snapshot of state obtained from a numerical solution to Eq.~\eqref{eq:final_derived_equ_dimless} at large time scales.
}
 \end{figure}

Colony formation is a pervasive phenomenon in living systems and
 is crucial for the survival of many species~\cite{Ben-Jacob_review_1998,Lejeune_2002,Higashi_2007,Taktikos_2014,biofilm_review_2014,PhysRevLett.79.313}.
 One of the well-known examples where colony formation is essential are biofilms.
  A bacterial colony can grow from a single cell via multiple cell divisions~\cite{Ben-Jacob_review_1998, biofilm_review_2014}. 
  However, there is another mechanism, which relies on successive encounters of individual, motile bacteria, as also occurring in the initial stages of biofilm formation.
This scenario of a kinetic formation of colonies dominates over proliferation if individuals are highly motile and their encounters drive the assembly of cells on a time scales much shorter than the characteristic cell division time. 
\emph{N. gonorrhoeae} or \emph{N. meningitidis} on biotic or abiotic substrates such as glass~\cite{Merz_nature_2000}, plastic (Fig.~\ref{fig:snapshots}(a)) or epithelial tissue~\cite{Higashi_2007} 
 are prototypical examples for such a scenario. 
 Motility of these and many other bacteria originates from long and thin filaments, called pili, 
 which grow out the cell, attach to a substrate, retract and thereby actively pull the cell forward~\cite{Rahul_Klumpp_2014, vasily_dave_2014,Maier_2002, Maier_review_2013}.
Pili are also used to mediate attractive displacements between cells~\cite{Merz_nature_2000,Craig_review_2004,Biais_2008, Maier_review_2013} with a characteristic interaction scale given by the mean pili length. 
Colonies begin to form within thirty minutes,
which is significantly smaller than the characteristic cell division time-scale (\emph{N. gonorrhoeae}: approx. $3$ h~\cite{Westling-Haeggstroem_1977}). 
Bacterial colonies are in general reversible structures.  Under certain conditions, for example the lack of nutrients or oxygen, they can dissolve and re-colonize their surroundings~\cite{Kolodkin-Gal30042010,Chamot-Rooke11022011,Gcell_dissolution_Maier_2015}. 
Specifically, \emph{N. meningitidis} and  \emph{N. gonorrhoeae} bacterial colonies have been shown to dissolve by effectively lowering the strength of the pili-mediated interaction~\cite{Chamot-Rooke11022011,Gcell_dissolution_Maier_2015}.

However, so far, the physical mechanisms governing the formation and dissolution of bacterial colonies are poorly understood. 
Since motility and interactions are driven by active retractions of pili,
fundamental concepts from equilibrium statistical mechanics are in general not applicable.
The inherent non-equilibrium nature of this system suggests to consider a  kinetic approach reminiscent of the Boltzmann equation, which has been successfully employed to describe the order-disorder transitions in several active systems far from equilibrium~\cite{Aronson_MT,Bertin_short,Bertin_long,Saintillan_2007,Aranson_Bakterien,Saintillan_2008,Ihle_2011,Thuroff_2013,Weber_NJP_2013,Hanke_2013,prx_with_flo_2014}. 

 Here we propose a kinetic description 
as a general framework of how living colonies form and dissolve, 
which keeps track of the length scales and the specific properties of the interactions between individuals. 
By a coarse-graining procedure we derive the corresponding hydrodynamic equation and find
an ordering instability for a choice of parameters relevant to \emph{N. gonorrhoeae}.
It belongs to a class of instabilities, where the diffusion constant is negative
and originates from attractive pili-mediated interactions.
As most of the parameters can be estimated based on available data for \emph{N. gonorrhoeae},
we analytically compute the corresponding phase-diagram
and the characteristic colony size that is consistent with experimental observations.

Our theory can also be used to compare the effects of different cell-cell interactions and investigate their interplay. 
We show that pili interactions are more effective regarding clustering than cell adhesion.
 Moreover, when both interactions keep the cells together in the colony, a more efficient and robust way to dissolve the colony is to lower the strength of pili-mediated interactions. This suggests that pili play an essential role not only in cell motility and assembly, but also in the dissolution of matured colonies. 
Our results demonstrate that kinetic theory can be applied to quantify the process of colony formation in living systems and is able to provide insights about the underlying
physical mechanisms.

\emph{Kinetic Model}: Our kinetic description is formulated in terms of the particle density $f(\mathbf{r},t)$.
We restrict ourselves to two-dimensional colonies forming on a planar substrate~\footnote{Since we focus on the onset of the instability, three-dimensional growth should be of secondary importance.}, which do not give rise to swarms or swirls~(see e.g.\ \cite{Aranson_Bakterien}).
Therefore, the spatial coordinates $\vec r\in\mathbb{R}^2$ suffice as dynamical variables.
 In the absence of interactions  cells are assumed to move across the substrate by pili-mediated displacements as in case of \emph{N. gonorrhoeae} or \emph{N. meningitidis}, leading to a diffusive behaviour at large length and time-scales~\cite{Holz_2010}. 
Interactions enter the kinetic description via ``collision rules".
 A collision rule $\mathcal{R}$ maps the pre-collision coordinates to the post-collision positions by means of the delta-functions $\delta(\cdot)$. 
The corresponding kinetic equation is: 
\begin{subequations}\label{eq:kinetic_equation_all}
\begin{equation}\label{eq:kinetic_equation}
	\partial_t f (\vec r,  t)= \mathcal{C}_\text{mot}(\vec r,  t) + \mathcal{C}_\text{int}(\vec r,  t),
\end{equation}
where $\mathcal{C}_\text{mot}$ describes the cell motility across the substrate
\begin{equation}\label{eq:single_cell_term}
	\mathcal{C}_\text{mot}(\vec r,  t)= \int \text{d}\vec r^\prime
	   \left[\mathcal{K}_{\vec r^\prime \to \vec r}  f(\vec r^\prime,  t) -  \mathcal{K}_{\vec r \to \vec r^\prime}  f(\vec r,  t) \right]
\end{equation}
and $\mathcal{C}_\text{int}$ accounts for the cell-cell interactions 
	\begin{eqnarray}	\label{eq:collision_term}
	& \mathcal{C}_\text{int}&(\vec r,  t)=\frac{1}{2}  \int \text{d}\vec r_1 \int \text{d}\vec r_2 \, \mathcal{W}(|\vec r_{12}|)  f(\vec r_1,  t) f(\vec r_2,  t)  \\
\nonumber
& \times& \left[  \delta \left(\mathcal{R}_1(\vec r_1,\vec r_2)
-\vec r \right)  + \delta \left(\mathcal{R}_2\left(\vec r_1,\vec r_2\right)-\vec r \right) - 2 \delta \left(\vec r_2-\vec r \right) \right].
\end{eqnarray}
$\mathcal{K}_{\vec r \to \vec r^\prime}$ denotes the transition kernel to move from $\vec r$ to $\vec r^\prime$ by a retraction event of an individual pilus. We assume that retraction events are independent and that the corresponding rate is isotropic, with a characteristic length scale  given by the pili length $\ell_\text{pi}$. There is  experimental evidence that the pili lengths are distributed exponentially~\cite{Holz_2010}. Therefore, we consider for the transition kernel  $\mathcal{K}_{\vec r \to \vec r + \vec{b}}=  \mathcal{K}_0 /(2\pi\ell_\text{pi}^2)  \exp{\left(-|\vec b|/\ell_\text{pi}\right)}$, with $ \mathcal{K}_0$ denoting the attachment rate of pili to the substrate and $\vec b=\vec r^\prime - \vec r$ is the displacement resulting from an individual pilus retraction.

$\mathcal{W}(|\vec r_{12}|)$ characterizes the isotropic kernel for collisions between cells with $|\vec r_{12}|=|\vec r_1- \vec r_2|$ denoting the relative cell-cell distance. 
 For pili-mediated attractive displacements, we consider the following collision rule:  
\begin{equation}\label{eq:collision_rule_one_half}
\left(\vec r_1,\vec r_2\right)  \rightarrow 
	\left(\mathcal{R}_1,\mathcal{R}_2\right)
	=\left(\vec r_1 - a \vec r_{12}, \vec r_2 +a \vec r_{12}\right),
\end{equation}
\end{subequations}
where $a\in[0,1/2]$ is a measure for the strength of the attractive interaction. 
For $a=1/2$, cells are maximally attracted and displaced to the center-of-mass coordinate $\vec R_{12}=\left(\vec r_1+\vec r_2\right)/2$ between the collision partners, 
 while for $a=0$, cells diffuse freely without interacting. 
Due to the exponential distribution of the pili lengths, the interaction rate is 
$\mathcal{W}_\text{pi}(|\vec r_{12}|) = \gamma \mathcal{W}_0/(2\pi\ell_\text{pi}^2)  \exp{\left(-|\vec r_{12}|/\ell_\text{pi}\right)}$, where $\ell_\text{pi}$
sets the characteristic length scale for the attractive interaction and $\mathcal{W}_0$ denotes the interaction rate.
Since pili-mediated cell-cell interactions are intrinsically stochastic~\cite{Rahul_Klumpp_2014,vasily_dave_2014},
we introduce a non-dimensional number, $\gamma$, accounting for the number of successful binding and retraction events to the total number of pili-cell encounter events. 
 
\emph{Coarse-graining}:  The isotropy of the interaction rates allows us to integrate Eq.~\eqref{eq:kinetic_equation_all} over the center-of-mass coordinates $\vec R_{12}$ leading to non-local terms (see Supplemental Material~\cite{SM},~S1).
 These terms are related to the length scales of the interactions  and resemble a phenomenological description for the assembly of active bundles~\cite{Kruse_2000, Kruse_2003a, Kruse_2003b}.
Since cell colonies typically exhibit sizes noticeably beyond the interaction length scale, 
the non-local integrands can be removed by expanding the particle density $f$ with respect to the spatial coordinates~\cite{Aronson_MT,Aranson_Bakterien}. Truncation of this expansion amounts to coarse-graining  beyond the interaction length scale.
To obtain a well-defined  set of hydrodynamic equations for the dynamics of bacterial colonies with pili-mediated interactions
we truncate at the fourth order (see Supplemental Material~\cite{SM},~S2):
\begin{eqnarray}\label{eq:final_derived_equ_dimless}
	\nonumber
	\partial_t \rho(\vec r,  t)&=&\alpha(\rho)   \nabla^2 \rho(\vec r,  t) -\beta_1 \, |\nabla\rho(\vec r,  t)|^2
	+\kappa(\rho) \, \nabla^4 \rho(\vec r,  t)  \\
	&+& \beta_2   \, \left[\nabla^2 \rho(\vec r,  t) \right]^2
	- \beta_3\, \left[\nabla \rho(\vec r,  t) \right] \cdot \nabla^3 \rho(\vec r,  t) ,
\end{eqnarray}
where  $\rho=f\cdot \ell_\text{pi}^2$ is the dimensionless density and the kinetic coefficients are
\begin{eqnarray}
\label{eq:kinetic_coeff_alpha}
\alpha(\rho)&=& G 
	-  \beta_1  \, \rho(\vec r,  t) , \\
 \label{eq:kinetic_coeff_kappa}
 \kappa(\rho)&=& -\left( \beta_2 +\beta_3 \right)
 \, \rho(\vec r,  t) ,
 \end{eqnarray}
 $\beta_1= a\bar a \tilde c_{2}$,  $\beta_2 = a^2 \bar a^2 \tilde c_{4}/4$, $\beta_3 =  \left(a\bar a^3+\bar a a^3\right) \tilde c_{4}/6$; $\bar a=1-a$. Note that all $\beta_i>0$.
 The  numerical constants $\tilde c_k$  are given in the Table of the Supplemental Material~\cite{SM}.
In Eq.~\eqref{eq:final_derived_equ_dimless}, we rescaled
 coordinates by the pili length 
 $\ell_\text{pi}$, i.e.\ $\vec r\to\vec r \cdot \ell_\text{pi}$, leading to a  rescaling of time \\
  $t\to t \cdot \ell_\text{pi}^2/(\mathcal{W}_0 \gamma)$.
We introduce the dimensionless parameter,
	$G=D/(\gamma \mathcal{W}_0)$, with $D=3\mathcal{K}_{0} \ell^2_\text{pi}$ denoting the single cell diffusion constant. 
$G$ is reminiscent of the inverse P\'eclet number and can be interpreted as a measure for the rate of diffusive particle transport relative to the frequency of interactions.
In other words, given a time period between two successive collisions, $G$ quantifies how much distance is traveled (on average) by diffusion with respect to the mean free path.

A  equation similar to Eq.~\eqref{eq:final_derived_equ_dimless}  but phenomenologically constructed appeared in the context of laminar flames
 and propagation of concentration waves referred to as Kuramoto-Sivashinsky equation~\cite{Kuramoto01021976,Sivashinsky}.
 It has also been pointed out as an
 appropriate framework to study instabilities in growing yeast colonies~\cite{PhysRevLett.79.313}. 
 However, Eq.~\eqref{eq:final_derived_equ_dimless} is distinctively different because the kinetic coefficients depend on density [Eqs.~\eqref{eq:kinetic_coeff_alpha} and \eqref{eq:kinetic_coeff_kappa}]. 
 Moreover, Eq.~\eqref{eq:final_derived_equ_dimless}  exhibits an alleged similarity to the Cahn-Hilliard equation studied in the context of liquid-liquid demixing~\cite{Bray_Review_1994}. 
Though both equations have terms of similar orders in $\mathcal{O}(\nabla \rho)$, they are fundamentally different with respect to the saturation of droplet or colony growth. 
The Cahn-Hilliard equation exhibits an instability of the homogeneous state, which saturates because the effective diffusion constant in front of the LaPlace operator decreases to zero.
Eq.~\eqref{eq:final_derived_equ_dimless} also exhibits an instability but it saturates due to a different mechanism as discussed below.

\begin{figure}[t]
\centering
\begin{tabular}{cc}
\includegraphics[width=0.25\textwidth]{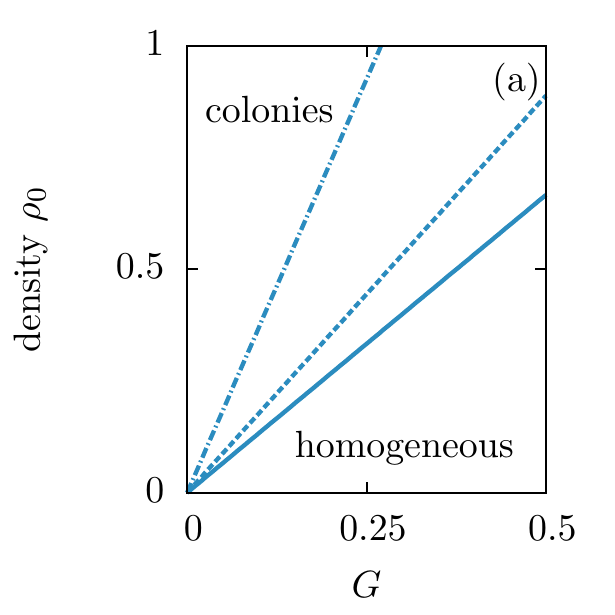} &
\includegraphics[width=0.25\textwidth]{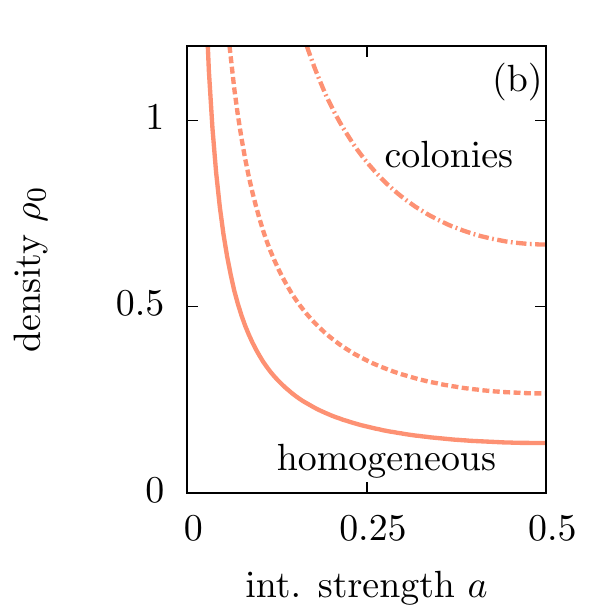} \\
\includegraphics[width=0.25\textwidth]{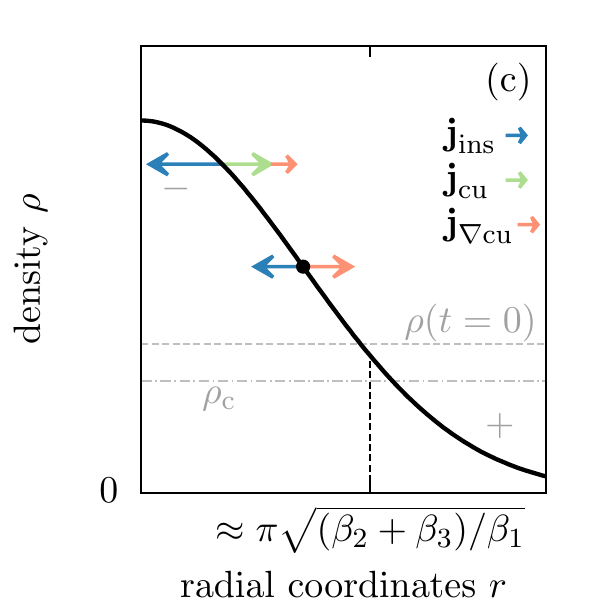}  &
\includegraphics[width=0.25\textwidth]{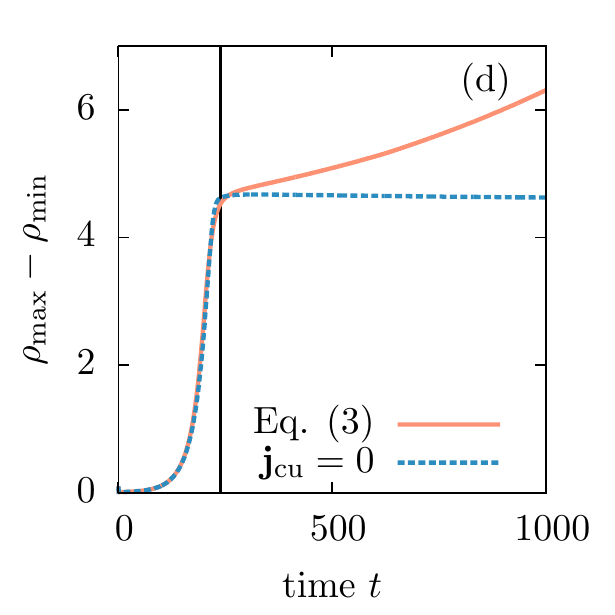}
\end{tabular}
\caption{\label{fig:phase_diagram} (color online) 
(a,b) Critical density $\rho_\text{c}$ as a function of the non-dimensional parameter $G$ and interaction strength $a$. 
Each line separates the parameter space, where colonies develop or the system remains homogeneous.
In \text{(b)}, three values of interaction strength $a$ are displayed: $(0.5, 0.25, 0.1)$ =(solid, dashed, dash-dotted),  and (c) depicts three values of $G$:
$(0.1, 0.2, 0.5)$=(solid, dashed, dash-dotted).
(c) Illustration of how the instability is balanced: For initial densities $\rho(t=0)>\rho_\text{c}$, the `instability flux' $\vec j_\text{ins}$ drives the emergence of 
a spatially inhomogeneous density profile. Depending on the location along the
density profile, the `curvature flux' $\vec j_\text{cu}$ and/or the `gradient curvature flux' $\vec j_{\nabla\text{cu}}$ acts against the `instability flux' $\vec j_\text{ins}$ and thereby balances the instability.
(d) Maximal density minus minimal density, $\rho_\text{max}-\rho_\text{min}$, as a function of time $t$, where  $\rho_\text{max}(t)=\text{max}_{\vec r} \rho(\vec r, t)$
and $\rho_\text{min}(t)=\text{min}_{\vec r} \rho(\vec r, t)$, for numerical solutions to Eq.~\eqref{eq:final_derived_equ_dimless} with and without `curvature flux' $\vec j_\text{cu}$. 
}
\end{figure}

\emph{Colony formation due to pili-mediated interaction:}  
The condition for the instability in Eq.~\eqref{eq:final_derived_equ_dimless}  is $\alpha(\rho)<0$.
Its onset marks 
 a critical density, $\rho_\text{c}=G/\beta_1$.
For $\rho_0> \rho_\text{c}$, the homogenous state of density $\rho_0$ is unstable.  
The instability enhances small density modulations around the homogenous density $\rho_0$ with a dispersion relation $w(q)= -\alpha(\rho_0) q^2 - \kappa(\rho_0) q^4$. 
$\rho_\text{c}$ depends on the non-dimensional parameter $G$ and the interaction strength $a$, $\rho_\text{c}=G/(a\bar a \tilde c_{2})$. We find that $\rho_\text{c}$ decreases for stronger attractive interactions, \ $a\to1/2$, and smaller values of $G$; see Fig.~\ref{fig:phase_diagram}(a,b).

The instability is opposed by fluxes related to the spatial curvature of the density field, which can be qualitatively understood by splitting the flux,
$\vec j = \vec j_\text{inst} +\vec j_\text{cu} + \vec j_{\nabla\text{cu}}$ with $\partial_t\rho=-\nabla\cdot \vec j$.
$\vec j_\text{inst}=-\alpha(\rho) \nabla \rho$ denotes the `instability flux' which acts for $\alpha<0$ like negative diffusion thus driving particles to the center of a density spot [see Fig.~\ref{fig:phase_diagram}(c) for an illustration].
There the instability current is opposed by the `curvature flux', $\vec j_\text{cu}= -\beta_2 (\nabla^2 \rho) \nabla \rho$, and the `gradient-curvature flux', $\vec j_{\nabla\text{cu}}= (\beta_2+\beta_3)  \rho \nabla ( \nabla^2 \rho)$. 
Both are directed outwards of the density spot since curvature is negative and increases.

Our findings on the instability and its saturation can be scrutinized by numerically solving  Eq.~\eqref{eq:final_derived_equ_dimless}. 
A representative snapshot of a state at large time-scales is shown in Fig.~\ref{fig:snapshots}(b), which appears to be similar to \emph{N. gonorrhoeae}  colonies  three hours after sedimentation on a plastic substrate [Fig.~\ref{fig:snapshots}(a)].
Using parameter values consistent with the experimental system we observe multiple colonies developing quickly for densities above the critical value.
We checked numerically that for all parameter values lying within the `colony phase' of the analytic phase diagram [Fig.~\ref{fig:phase_diagram}(a,b)] give rise to the formation of colonies.
After the onset of the instability, colonies exponentially grow with a growth speed that is higher the larger  the difference of the homogeneous density $\rho_0$ to the critical density $\rho_\text{c}$; see Supplemental Material \cite{SM}, S.4. 
Thus, for   $\rho_0 \searrow^+ \rho_\text{c}$, we observe a colony growth rate decreasing to zero; a phenomena reminiscent of `critical slowing down' in phase transitions~\cite{Onuki_book}.
Subsequent to the initial growth, there is regime, where colonies  grow only very slowly [Fig.~\ref{fig:phase_diagram}(d), solid red line]. 
The later observation is  due to a weak interaction between the colonies via some evaporation-condensation mechanism qualitatively reminiscent  of Ostwald-ripening in liquid-liquid phase-separation~\cite{Bray_Review_1994}. 
Interestingly, at the onset of the instability the non-linear `curvature flux' $\vec j_\text{cu}$ vanishes
suggesting that it  might play an essential role for developed colonies at large time-scales.
Running the system without curvature flux, $\vec j_\text{cu}= 0$, we find that the subsequent ripening is absent leading to a stable state consisting of multiple colonies [see Fig.~\ref{fig:phase_diagram}(d), dashed line]. 
This implies that interactions between colonies is driven by the `curvature flux', while the `gradient curvature flux' suffices for the saturation. 
Based on this insight we can analytically estimate the colony size at the time when the system 
crosses to the very slow ripening regime [vertical line in Fig.~\ref{fig:phase_diagram}(d)] by neglecting the curvature flux (see Supplemental Material \cite{SM}, S.3 for more details).
 For densities $\rho \gg \rho_\text{c}$,
stationary periodic solutions are supported with 
 the quasi-static colony size of  $\pi \sqrt{(\beta_2+\beta_3)/\beta_1}$.
 Remarkably, for $a=0.5$,
 this estimate suggests a colony size of several pili-lengths, which is consistent with 
\emph{N. gonorrhoeae} [Fig.~\ref{fig:snapshots}(a)].

\emph{Biological relevance}: 
In principle, all parameters entering the kinetic description Eq.~\eqref{eq:kinetic_equation_all}  can be measured or estimated for living colonies forming on a substrate and thereby all kinetic coefficients in Eq.~\eqref{eq:final_derived_equ_dimless}. 
In particular, for \emph{N. gonorrhoeae},
$\ell_\text{pi}\approx 1 \mu m$~\cite{Holz_2010,vasily_dave_2014}
and colony formation is observed for densities of $\rho \approx 0.2$. 
The  attachment rate to the substrate can be obtained from measurements of the single cell diffusion constant, $\mathcal{K}_0=D/(3\ell_\text{pi}^2)\approx (6s)^{-1}$ with $D\approx 0.5 \mu m^2/s$~\cite{Taktikos_2014} and the cell-cell interaction rate can be roughly estimated from the experimental value of the mean next neighbour distance and the mean pili-number per cell  
to $\mathcal{W}_0/\ell_\text{pi}^2 \sim 5 s^{-1}$ (see Supplemental Material~\cite{SM}, S7). Therefore, a typical value for the dimensionless 
parameter for \emph{N. gonorrhoeae} is $G\sim 0.1 \gamma^{-1}$.
Recently, the  attachment probability of pili to a substrate has been determined by fitting a model to experimental results~\cite{vasily_dave_2014}, finding an approximate value of $0.5$. We expect a roughly similar, maybe lower value for $\gamma$ since successful binding to another cell can be hindered
by other moving cells.
So far an appropriate estimate for the interaction strength $a$ is missing  
because the synchronous  visualization of pili and cell movement is not feasible for large enough time-scales.  Thereby, we consider $a$ as an unknown parameter. 

The proposed kinetic description, Eq.~\eqref{eq:kinetic_equation_all},
can also be used to include other attractive interactions such as adhesion.
Since cell-cell adhesion constitutes a local interaction on the scale of the cell diameter, an appropriate weight function is for example a Gaussian of the form
$\mathcal{W}_\text{ad}(|\vec r_{12}|/\ell_\text{ad}) =  \mathcal{W}_0/(\pi \ell_\text{ad}^2)  \exp{\left(-\vec r_{12}^2/\ell_\text{ad}^2\right)},$
where $\ell_\text{ad}$ denotes the characteristic length scale which is in the order of the cell size.
Comparing both interactions (see Supplemental Material for details~\cite{SM}, S.5) we find that  pili allow for a significantly more pronounced affinity for colony formation compared to adhesive interactions, i.e.\ colonies already form at smaller initial density of cells.

\begin{figure}[t!]
\centering
\begin{tabular}{cc}
\includegraphics[width=0.25\textwidth]{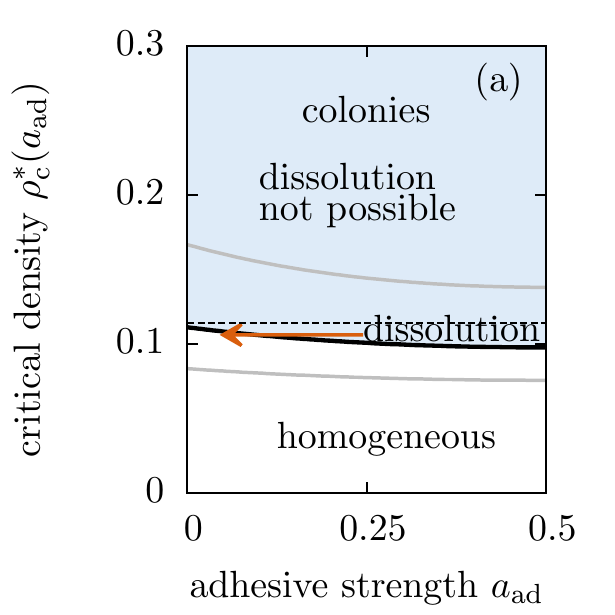} & 
\includegraphics[width=0.25\textwidth]{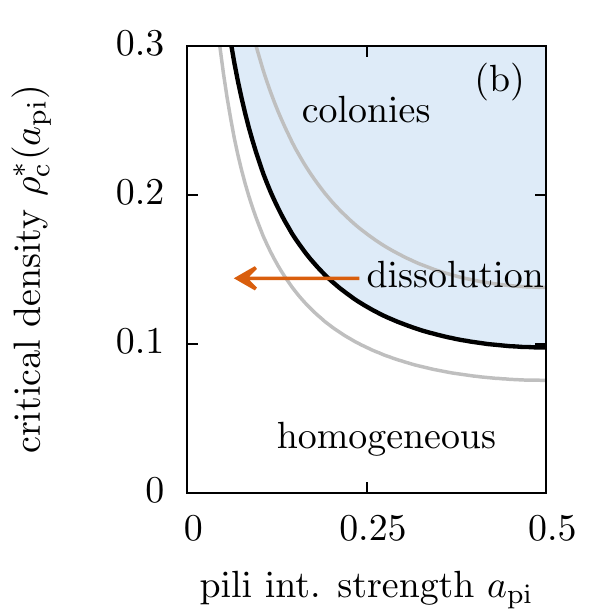}\\
\end{tabular}
\caption{\label{fig:numerics_and_dissolution} 
(color online) 
Critical density  $\rho^*_\text{c}(a_\text{ad},a_\text{pi})/G$ 
as a function of  \text{(a)} adhesive strength $a_\text{ad}$ (with $a_\text{pi}=0.5$) and  \text{(b)} pili-mediated interaction strength  $a_\text{pi}$ (with $a_\text{ad}=0.5$). In both plots, $G=0.1$ and each solid line  corresponds to $\gamma\in\{0.4, 0.3,0.2\}$ from top to bottom.
Black line corresponds to $\gamma=0.3$. 
The horizontal black dashed line (plotted only for $\gamma=0.3$ in (a)) marks the dissolution boundary: Below, dissolution (red arrow) is possible, above not. Blue shaded areas  correspond to the ``colony-phase".
} 
\end{figure} 

 \emph{Colony dissolution:} Many bacteria are known to interact simultaneously by adhesion and pili. It is hypothesized that  these bacteria are able to switch off either adhesion or the pili-mediated interaction without affecting their ability to move ~\cite{Chamot-Rooke11022011,EMI:EMI775,  Gcell_dissolution_Maier_2015}.
 Now we address the question of whether developed colonies can dissolve by switching off either one of these interactions.
 In other words, given the phase-diagram of a specific bacteria system, 
 we discuss some possible means of leaving the ``colony-phase" by $a_\text{pi}\to 0$ or $a_\text{ad}\to 0$.
We now include both interactions
by adding a term for  pili-mediated interactions $\mathcal{C}_\text{int,pi}$ and a term corresponding to adhesive interactions $\mathcal{C}_\text{int,ad}$ on the right hand side of Eq.~\eqref{eq:kinetic_equation}, i.e.\ $\mathcal{C}_\text{int}=\mathcal{C}_\text{int,pi}+\mathcal{C}_\text{int,ad}$.
In addition to the already introduced different length scales $\ell_\text{pi}$ and $\ell_\text{ad}$, we also distinguish the corresponding interaction strengths, denoted as $a_\text{pi}$ and $a_\text{ad}$ (values for adhesion and pili-mediated interactions are denoted as $\tilde c_{k,\text{ad}}$ and $\tilde c_{k,\text{pi}}$). 
We rescale  coordinates,  density and time by the adhesive interaction length $\ell_\text{ad}$ (or cell size), i.e.\ $\vec r\to\vec r \cdot\ell_\text{ad}$, $f\to f / \ell_\text{ad}^2\equiv \rho$ and $t\to t \cdot \ell_\text{ad}^2/\mathcal{W}_0$, thereby introducing a ratio of these length scales,  $\epsilon=\ell_\text{pi}/\ell_\text{ad}$.
  For the case where cells interact with both adhesive and pili-mediated interactions,
 we find the following effective diffusion constant  (further coefficients see Supplemental Material~\cite{SM},~S6):
$\alpha(\rho)= G 
	- \rho \,\left[ a_\text{ad} \bar a_\text{ad}  \, \tilde c_{2,\text{ad}}  +  a_\text{pi} \bar a_\text{pi}  \,  \tilde c_{2,\text{pi}} \, \gamma \, \epsilon^2 \right]$.
Setting this equation equal to zero marks
 a critical density
$\rho^*_\text{c}(a_\text{ad}, a_\text{pi})= G /\left[ a_\text{ad} \bar a_\text{ad}  \, \tilde c_{2,\text{ad}}  + a_\text{pi} \bar a_\text{pi}   \,  \tilde c_{2,\text{pi}} \, \gamma \, \epsilon^2 \right]$ depending on the strength of both interactions,  $a_\text{ad}$ and $a_\text{pi}$. 

In order to study the impact of both interactions for dissolution of colonies we choose the parameters ($\epsilon$, $\gamma$) relevant to \emph{N. gonorrhoeae}.
Fig.~\ref{fig:numerics_and_dissolution}(a)  shows $\rho^*_\text{c}(a_\text{ad})$ as a function of $a_\text{ad}$ for $a_\text{pi}=0.5$ and $G=0.1$, while Fig.~\ref{fig:numerics_and_dissolution}(b) depicts $\rho^*_\text{c}(a_\text{pi})$ as a function of $a_\text{pi}$ for $a_\text{ad}=0.5$ and $G=0.5$, both  for several values of $\gamma$.
 For a given $\gamma$, there are two qualitatively distinct regimes for the case where adhesive interactions are switched off [Fig.~\ref{fig:numerics_and_dissolution}(a)]:
 For small enough $\rho^*_\text{c}$ below the ``dissolution boundary" (horizontal dashed line), colonies can dissolve by switching off the adhesive interaction ($a_\text{ad}\to 0$) and is indicated by the red arrows.
 However, above the dissolution boundary, colonies cannot dissolve. 
Interestingly,  choosing the parameters relevant to \emph{N. gonorrhoeae} 
gives a rather small density regime, where colonies can dissolve,  rendering the 
 dissolution scenario through switching off adhesion as  a non-robust mechanism.
 This is in stark contrast to the scenario of switching off pili-mediated interactions [Fig.~\ref{fig:numerics_and_dissolution}(b)]:
For a given $\gamma$, dissolution is possible for all experimental densities in the ``colony-phase'' by lowering the pili-interaction strength, $a_\text{pi}\to 0$.
These findings suggest that switching off pili-mediated interactions is a more robust mechanism for the dissolution 
of bacterial colonies than switching off adhesion.

To summarize, the formation of living colonies is investigated using a hydrodynamic equation derived from a kinetic description, where most of the parameters can be estimated from experimental data for  \emph{N. gonorrhoeae} bacteria.
Our results demonstrate that kinetic theory can be successfully used to describe 
complex far from equilibrium systems such as formation and dissolution of living bacterial colonies.
Applications of this theory could pave the way for the physical quantification of the initial stages of
biofilm formation.  
Though biological reasons for colony formation are specific to each system 
there are qualitative similarities~\cite{Ben-Jacob_review_1998,Lejeune_2002,Higashi_2007,Taktikos_2014,biofilm_review_2014,PhysRevLett.79.313}: Colonies form due to encounters with nearby individuals giving rise to structures of a characteristic size determined by the intra-species interactions and the environment. 
These similarities suggest that our kinetic description might be applied to other colony-forming systems while the kinetic coefficients in the resulting hydrodynamic equation may differ for each system. 
Further open questions concern the role of cell division and stochastic fluctuations in living colonies~\cite{Tsimring_review_noise_biology_2014}.

\begin{acknowledgments}
We thank Igor S. Aranson, Frank J\"ulicher and Florian Th\"uroff for their very insightful comments on this manuscript, and  Coleman Broaddus for his careful revisions. 
We acknowledge funding from the NIH (grant AI116566 to N.B.).
\end{acknowledgments}

\newpage


\cleardoublepage
\onecolumngrid


\section*{Supplementary material}

# Supplemental Material:
# Formation and Dissolution of Bacterial Colonies

Christoph A. Weber,[1] Yen Ting Lin,[1] Nicolas Biais,[2] and Vasily Zaburdaev[1]

[1]*Max Planck Institute for the Physics of Complex Systems, Nöthnitzer Str. 38, 01187 Dresden, Germany*
[2]*Department of Biology, Brooklyn College, City University of New York, Brooklyn, NY 11210*

## S1. Coordinate change due to isotropy of interaction kernel

Using the collision rule (see Eq. (2), main text; and Fig. S1),

$$(\mathbf{r}_1, \mathbf{r}_2) \to (\mathcal{R}_1, \mathcal{R}_2) = (\mathbf{r}_1 - a\,\mathbf{r}_{12}, \mathbf{r}_2 + a\,\mathbf{r}_{12}), \quad \text{(S1)}$$

the gain term $\mathcal{C}^+$ can be written as:

$$\mathcal{C}^+ = \int d\mathbf{r}_1 \int d\mathbf{r}_2\, \mathcal{W}(\mathbf{r}_1, \mathbf{r}_2)\, f(\mathbf{r}_1, t) f(\mathbf{r}_2, t)$$
$$\times \frac{1}{2}\Big( \delta\left((\mathbf{r}_1 - a \cdot \mathbf{r}_{12}) - \mathbf{r}\right) + \delta\left((\mathbf{r}_2 + a \cdot \mathbf{r}_{12}) - \mathbf{r}\right) \Big). \quad \text{(S2)}$$

The equation above can be rewritten in terms of relative coordinates $\mathbf{r}_{12} = \mathbf{r}_1 - \mathbf{r}_2$ and center-of-mass coordinates $\mathbf{R}_{12} = (\mathbf{r}_1 + \mathbf{r}_2)/2$, i.e. $\mathbf{r}_1 = \mathbf{R}_{12} + \mathbf{r}_{12}/2$ and $\mathbf{r}_2 = \mathbf{R}_{12} - \mathbf{r}_{12}/2$ (see Fig. S1):

$$\mathcal{C}^+ = \frac{1}{2} \int d\mathbf{r}_{12} \int d\mathbf{R}_{12}\, \mathcal{W}(|\mathbf{r}_{12}|)$$
$$\times f(\mathbf{R}_{12} + \mathbf{r}_{12}/2, t) f(\mathbf{R}_{12} - \mathbf{r}_{12}/2, t)$$
$$\times \Big[ \delta\left(\mathbf{R}_{12} + \mathbf{r}_{12}\left(\frac{1}{2} - a\right) - \mathbf{r}\right)$$
$$+ \delta\left(\mathbf{R}_{12} + \mathbf{r}_{12}\left(-\frac{1}{2} + a\right) - \mathbf{r}\right) \Big]. \quad \text{(S3)}$$

Due to the isotropy of the collision kernel the integration over the center-of-mass coordinates can be performed, finding

$$\mathcal{C}^+ = \frac{1}{2} \int d\mathbf{r}_{12}\, \mathcal{W}(|\mathbf{r}_{12}|)\Big[ f(\mathbf{r} + a\mathbf{r}_{12}, t) f(\mathbf{r} - \bar{a}\mathbf{r}_{12}, t)$$
$$+ f(\mathbf{r} - a\mathbf{r}_{12}, t) f(\mathbf{r} + \bar{a}\mathbf{r}_{12}, t)\Big], \quad \text{(S4)}$$

where $\bar{a} = 1 - a$. The equation above can be further simplified for example for the case $a = \frac{1}{2}$:

$$\mathcal{C}^+ = \int d\mathbf{r}_{12}\, \mathcal{W}(|\mathbf{r}_{12}|) f(\mathbf{r} + \mathbf{r}_{12}/2, t) f(\mathbf{r} - \mathbf{r}_{12}/2, t). \quad \text{(S5)}$$

Similar manipulations can be performed for the loss term:

$$\mathcal{C}^- = \int d\mathbf{r}_1 \int d\mathbf{r}_2\, \mathcal{W}(|\mathbf{r}_{12}|)\, f(\mathbf{r}_1, t) f(\mathbf{r}_2, t)\, \delta(\mathbf{r}_2 - \mathbf{r})$$
$$= \int d\mathbf{r}_{12} \int d\mathbf{R}_{12}\, \mathcal{W}(|\mathbf{r}_{12}|)\, \delta(\mathbf{R}_{12} - (\mathbf{r}_{12}/2 + \mathbf{r}))$$
$$\times f(\mathbf{R}_{12} + \mathbf{r}_{12}/2, t) f(\mathbf{R}_{12} - \mathbf{r}_{12}/2, t)$$
$$= f(\mathbf{r}, t) \int d\mathbf{r}_{12}\, \mathcal{W}(|\mathbf{r}_{12}|)\, f(\mathbf{r} + \mathbf{r}_{12})$$
$$\equiv f(\mathbf{r}, t) \int d\mathbf{r}_{12}\, \mathcal{W}(|\mathbf{r}_{12}|) f(\mathbf{r} - \mathbf{r}_{12}). \quad \text{(S6)}$$

Please note that non-local integrands above resemble a phenomenological description for the assembly of active bundles [1–3].

## S2. Details of coarse-graining and truncation

Truncating of the non-local distribution function $f(\mathbf{r} \pm a\mathbf{r}_{12})$ (cf. Eq. (S4) and Eq. (S6)) at the fourth order leads to

$$f(\mathbf{r} \pm a\mathbf{r}_{12}) = f(\mathbf{r}) \pm a\left(\mathbf{r}_{12} \cdot \nabla\right) f(\mathbf{r}) + \frac{a^2}{2}\left(\mathbf{r}_{12} \cdot \nabla\right)^2 f(\mathbf{r})$$
$$\pm \frac{a^3}{6}\left(\mathbf{r}_{12} \cdot \nabla\right)^3 f(\mathbf{r}) + \frac{a^4}{24}\left(\mathbf{r}_{12} \cdot \nabla\right)^4 f(\mathbf{r}) + \mathcal{O}\left[\left(\mathbf{r}_{12} \cdot \nabla\right)^5\right], \quad \text{(S7)}$$

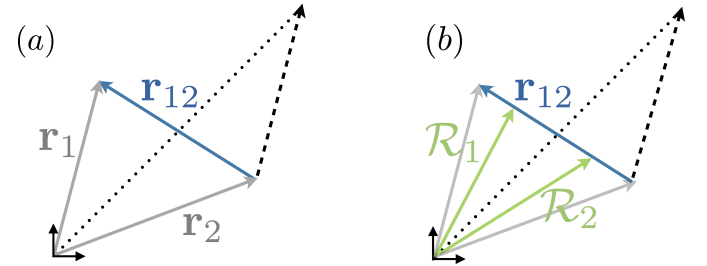

FIG. S1. Illustration of collision setup and the collision rule mimicking attractive interactions. **(a)** The two collision partners have the spatial coordinates $\mathbf{r}_1$ and $\mathbf{r}_2$, defining a relative distance $\mathbf{r}_{12} = \mathbf{r}_1 - \mathbf{r}_2$ and a center-of-mass coordinate $\mathbf{R}_{12} = (\mathbf{r}_1 + \mathbf{r}_2)/2$. **(b)** The collision given in Eq. (2) [main text], which maps the pre-collision coordinates $(\mathbf{r}_1, \mathbf{r}_2)$ to the post-collision coordinates $(\mathcal{R}_1(\mathbf{r}_1, \mathbf{r}_2), \mathcal{R}_2(\mathbf{r}_1, \mathbf{r}_2))$ [indicated by green arrows]. For the illustration, $a = 0.25$.

where we omitted the time dependence for reasons of brevity. Using the truncation above and neglecting all terms $\mathcal{O}\left[(\mathbf{r}_{12}\cdot\nabla)^5\right]$ amounts to an explicit coarse-graining of the system's dynamics to length scales beyond the characteristic length scale of the interaction. Therefore, we will refer to the resulting equation obtained at the end of this section as hydrodynamic equation.

### 1. Single cell motility term:

The term modeling the single cell motility across the substrate, $\mathcal{C}_{\text{mot}}$, can be coarse-grained as follows. Defining the pili-mediated displacement as $\mathbf{b} = (b_x, b_y) = \mathbf{r}' - \mathbf{r}$, and the transition rate $\mathcal{K}_{\mathbf{r}\to\mathbf{r}'} \equiv \mathcal{K}(\mathbf{b}; \mathbf{r})$:

$$\mathcal{C}_{\text{mot}}(\mathbf{r}, t) = \int d\mathbf{r}' \left[\mathcal{K}_{\mathbf{r}'\to\mathbf{r}} f(\mathbf{r}', t) - \mathcal{K}_{\mathbf{r}\to\mathbf{r}'} f(\mathbf{r}, t)\right]$$
$$= \int d\mathbf{b} \left[\mathcal{K}(-\mathbf{b}; \mathbf{r} + \mathbf{b}) f(\mathbf{r}', t) - \mathcal{K}(\mathbf{b}; \mathbf{r}) f(\mathbf{r}, t)\right]$$
$$= \int d\mathbf{b} \left[\mathcal{K}(\mathbf{b}; \mathbf{r} - \mathbf{b}) f(\mathbf{r} - \mathbf{b}, t) - \mathcal{K}(\mathbf{b}; \mathbf{r}) f(\mathbf{r}, t)\right]. \quad (S8)$$

Expanding the non-local integrand with respect to the spatial coordinates and keeping only the highest non-vanishing order leads to:

$$\mathcal{C}_{\text{mot}}(\mathbf{r}, t) = \int d\mathbf{b} \frac{1}{2} (\mathbf{b}\cdot\nabla)^2 \left[\mathcal{K}(\mathbf{b}; \mathbf{r}) f(\mathbf{r}, t)\right]$$
$$= \int d\mathbf{b} \frac{1}{2} \mathcal{K}(\mathbf{b}) \left(b_x \partial_x + b_y \partial_y\right)^2 f(\mathbf{r}, t)$$
$$= \int d\mathbf{b} \frac{1}{2} \mathcal{K}(\mathbf{b}) \left(b_x^2 \partial_x^2 + b_y^2 \partial_y^2\right) f(\mathbf{r}, t) \quad (S9)$$
$$= \left[\int d\mathbf{b} \mathcal{K}(|\mathbf{b}|) b_x^2\right] \left(\partial_x^2 + \partial_y^2\right) f(\mathbf{r}, t)$$
$$= D\nabla^2 f(\mathbf{r}, t),$$

where we assumed that the transition rate does not depend on the spatial coordinates, $\mathcal{K}(\mathbf{b}; \mathbf{r}) = \mathcal{K}(\mathbf{b})$, and that it is an even [4] and isotropic function, $\mathcal{K}(\mathbf{b}) = \mathcal{K}(|\mathbf{b}|)$. Moreover, using $\mathcal{K}(|\mathbf{b}|) = \mathcal{K}_0 \frac{1}{2\pi\ell_{\text{pi}}^2} \exp(-|\mathbf{b}|/\ell_{\text{pi}})$ one obtains the diffusion constant for pili-mediated motility:

$$D = \mathcal{K}_0 \ell_{\text{pi}}^2 \tilde{c}_{2,\text{pi}}, \quad (S10)$$

where $\tilde{c}_{2,\text{pi}}$ given in Table S1 and $\mathcal{K}_0$ denotes the attachment rate of pili to the substrate. Restriction to the second order is validated by the *Pawula* theorem [5].

### 2. Interaction term:

The interaction term $\mathcal{C}_{\text{int}}$ can be split in a gain term, $\mathcal{C}^+$, and a loss term, $\mathcal{C}^-$.

*Gain:* Neglecting all terms above the fourth order, we find for the loss term:

$$\mathcal{C}^- = f(\mathbf{r})\left[c_0 f(\mathbf{r}) + \frac{c_2}{2}\nabla^2 f(\mathbf{r}) + \frac{c_4}{24}\nabla^4 f(\mathbf{r})\right], \quad (S11)$$

where the coefficients are given as

$$c_k = \int\int dr_{12,x} dr_{12,y} \; r_{12,x}^k \; \mathcal{W}_{\text{ad/pi}}(r_{12})$$
$$= \mathcal{W}_0 \begin{cases} \ell_{\text{ad/pi}}^k \cdot \tilde{c}_k & \text{if } k \text{ even,} \\ 0 & \text{if } k \text{ odd,} \end{cases} \quad (S12)$$

with $r_{12} = |\mathbf{r}_{12}| = \sqrt{r_{12,x}^2 + r_{12,y}^2}$ and $\tilde{c}_k$ are dimensionless numbers given in Table S1 for the adhesive(ad) and pili-mediated(pi) interaction. Note that odd powers, thereby also mixed gradients such as $\partial_x \partial_y f$, vanish since the integral is then an asymmetric function with respect to the integration over e.g. $dr_{12,x}$ or $dr_{12,x}$, respectively [6].

*Loss:* Following similar lines for the source term $\mathcal{C}^+$, and neglecting all contributions above the fourth order in the spatial derivatives, we find only six non-zero contributions:

$$\mathcal{C}_{\text{int}}^+ = c_0 f^2(\mathbf{r}) + \frac{1}{2}[a^2 + \bar{a}^2] c_2 f(\mathbf{r})\nabla^2 f(\mathbf{r}) - a\bar{a}c_2 |\nabla f(\mathbf{r})|^2$$
$$+ \frac{1}{24}[a^4 + \bar{a}^4] c_4 f(\mathbf{r})\nabla^4 f(\mathbf{r}) + \frac{1}{4}a^2\bar{a}^2 c_4 \left[\nabla^2 f(\mathbf{r})\right]^2$$
$$- \frac{1}{6}[a\bar{a}^3 + \bar{a}a^3] c_4 \left[\nabla f(\mathbf{r})\right]\cdot\nabla^3 f(\mathbf{r}). \quad (S13)$$

*Gain and loss:* Combining gain and loss term, $\mathcal{C}_{\text{int}} = \mathcal{C}^+ - \mathcal{C}^-$, and plugging it in Eq. (1) [main text], leads to the final hydrodynamic equation:

$$\partial_t f(\mathbf{r}, t) = D\nabla^2 f(\mathbf{r}, t) + \frac{1}{2}[a^2 + \bar{a}^2 - 1] c_2 f(\mathbf{r}, t)\nabla^2 f(\mathbf{r}, t)$$
$$+ \frac{1}{24}[a^4 + \bar{a}^4 - 1] c_4 f(\mathbf{r}, t)\nabla^4 f(\mathbf{r}, t)$$
$$- a\bar{a}c_2 |\nabla f(\mathbf{r}, t)|^2 + \frac{1}{4}a^2\bar{a}^2 c_4 \left[\nabla^2 f(\mathbf{r}, t)\right]^2$$
$$- \frac{1}{6}[a\bar{a}^3 + \bar{a}a^3] c_4 \left[\nabla f(\mathbf{r}, t)\right]\cdot\nabla^3 f(\mathbf{r}, t). \quad (S14)$$

Note that in Eq. (S14) the zeroth order term cancels because of particle conservation.

As last step, we use the scaling of the kinetic coefficients, $c_k = \mathcal{W}_0 \ell^k \tilde{c}_k$ [see Eq. (S12)], with $\ell \in \{\ell_{\text{ad}}, \ell_{\text{pi}}\}$ and the numerical values $\tilde{c}_k$ given in Table S1, and write Eq. (S14) in a dimensionless form. To this end, we rescale the coordinates and the density by means of the interaction length $\ell$, i.e. $\mathbf{r} \to \mathbf{r}\cdot\ell$ and $f \to f/\ell^2 \equiv \rho$. This implies a rescaling of the time-scale given by $t \to t\cdot\ell^2/(\mathcal{W}_0\gamma)$; note that $\gamma = 1$ for adhesive interactions. Along these lines we find the following dimensionless parameter

$$G = \frac{D}{\gamma\mathcal{W}_0}, \quad (S15)$$

| $k$ | 0 | 1 | 2 | 3 | 4 |
|---|---|---|---|---|---|
| $\tilde{c}_{k,\mathrm{ad}}$ | 1 | 0 | $\frac{1}{2}$ | 0 | $\frac{3}{4}$ |
| $\tilde{c}_{k,\mathrm{pi}}$ | 1 | 0 | 3 | 0 | 45 |

TABLE S1. The numerical numbers $\tilde{c}_k$ (as defined in Eq. (S12)) corresponding to adhesion and pili-mediated interactions (refer to main text for the respective collision kernels).

where $D$ denotes the single cell diffusion constant given by Eq. (S10).

Using the aforementioned rescalings leads to the following dimensionless equation:

$$\partial_t \rho(\mathbf{r},t) = \alpha(\rho)\nabla^2 \rho(\mathbf{r},t) + \kappa(\rho)\nabla^4\rho(\mathbf{r},t) - \beta_1\left|\nabla\rho(\mathbf{r},t)\right|^2 \\ + \beta_2\left[\nabla^2\rho(\mathbf{r},t)\right]^2 - \beta_3\left[\nabla\rho(\mathbf{r},t)\right]\cdot\nabla^3\rho(\mathbf{r},t) ,\tag{S16}$$

where the kinetic coefficients are defined as:

$$\alpha(\rho) = G - \beta_1\,\rho(\mathbf{r},t),\tag{S17}$$

$$\kappa(\rho) = -\left(\beta_2 + \beta_3\right)\,\rho(\mathbf{r},t),\tag{S18}$$

$$\beta_1 = a\bar{a}\tilde{c}_2,\tag{S19}$$

$$\beta_2 = \frac{1}{4}a^2\bar{a}^2\tilde{c}_4,\tag{S20}$$

$$\beta_3 = \frac{1}{6}\left(a\bar{a}^3 + \bar{a}a^3\right)\tilde{c}_4,\tag{S21}$$

and with numerical values $\tilde{c}_k$ given in Table S1 and $\bar{a} = 1 - a$. It is worth mentioning that Eq. (S16) conserves density, i.e. it can be written as $\partial_t\rho = -\nabla\cdot\mathbf{j}$, with the flux $\mathbf{j}$ given by [7]:

$$\mathbf{j} = -\alpha(\rho)\nabla\rho - \left[\kappa(\rho)\nabla^2 + \beta_2\left(\nabla^2\rho\right)\right]\nabla\rho.\tag{S22}$$

### S3. Quasi-stationary colony size

An approximative expression for the colony size when the system transits from fast initial growth to the slow ripening regime (see main text) can be obtained by neglecting the 'curvature flux', $\mathbf{j}_{\mathrm{cu}} = \beta_2\left(\nabla^2\rho\right)\nabla\rho$ in Eq. (S22). Note that the 'curvature flux' also vanishes after linearization of Eq. (S22) around $\rho_0$ with $\rho = \rho_0 + \delta\rho$. Assuming quasi-static conditions, $\mathbf{j} = 0$, leads to

$$0 = \left[\alpha(\rho_0) + \kappa(\rho_0)\nabla^2\right]\nabla\delta\rho.\tag{S23}$$

Within this quasi-static approximation there are two stationary states: The homogenous density field $\rho = \rho_0$ with $\nabla\rho = 0$ and an inhomogeneous state ($\nabla\rho \neq 0$) that supports periodic solutions suggesting the coexistence of several colonies. Since our simulations indicate that droplets have a very similar size when the system crosses from the fast initial growth to the slow ripening phase (see video material), let us approximate the droplet size distribution to be infinitely narrow and extract a single length

scale, referred to as quasi-stationary colony size $\xi$. Writing $\nabla \to i\mathbf{q}$ and $\xi = \pi|\mathbf{q}|^{-1}$, one finds for $\nabla\rho \neq 0$:

$$\xi^2(\rho_0) \simeq \pi^2\frac{\kappa(\rho_0)}{\alpha(\rho_0)}.\tag{S24}$$

For large densities, one gets:

$$\xi(\rho_0 \to \infty) \simeq \pi\sqrt{\frac{\beta_2 + \beta_3}{\beta_1}}\tag{S25}$$

which gives approximately 5 pili-length for $a = 0.5$; a value that is consistent with *N. gonorrhoeae* colonies forming on a substrate [see Fig. 1(a), main text].

### S4. Numerical solution

Eq. (S16) was solved numerically using a 8th (embedded 9th) order adaptive Runge-Kutta method. Specifically, we employed an established package called "XMDS2" (see [8]) developed by Graham R. Dennis, Joseph J. Hope and Mattias T. Johnsson (Comput. Phys. Commun. 184, 201–208 (2013)).

Fig. S2 depicts numerical solutions of Eq. (S16) for various density values in order to illustrate that the equation—originating from a gradient expansion and a truncation at the fourth order—is well-behaved in case of an instability. For a discussion of the results, please refer to the figure caption of Fig. S2.

### S5. Phase diagrams: Pili-mediated interaction versus adhesion

Since the pili length distribution exhibits a more pronounced tail than the localised Gaussian distribution (characterised by larger $\tilde{c}_2$) and also a larger characteristic length as for *N. gonorrhoeae* ($\ell_{\mathrm{pi}} > \ell_{\mathrm{ad}}$), pili allow for a significantly more pronounced affinity for colony formation compared to adhesive interactions [Fig. S3(a,b)].

### S6. Kinetic coefficients for adhesion and pili-mediated interactions

For adhesive and pili-mediated interactions, solely the effective diffusion constant $\alpha(\rho)$ has been given in the main text. Reading the coefficients Eq. (S19)–(S21) as function of the interaction strength $a$, i.e. $\beta_i = \beta_i(a)$, the remaining coefficient is listed below:

$$\kappa(\rho) = -\Big[\big(\beta_2(a_{\mathrm{ad}}) + \beta_3(a_{\mathrm{ad}})\big) \\ + \big(\beta_2(a_{\mathrm{pi}}) + \beta_3(a_{\mathrm{pi}})\big)\,\gamma\,\epsilon^4\Big].\tag{S26}$$

The corresponding length scales, interaction strength and numerical coefficients for adhesion and pili-mediated interactions are denoted as: $\ell_{\mathrm{pi}}, a_{\mathrm{ad}}, \tilde{c}_{k,\mathrm{ad}}$ and $\ell_{\mathrm{ad}}, a_{\mathrm{ad}}, \tilde{c}_{k,\mathrm{ad}}$, respectively, and $\epsilon = \ell_{\mathrm{pi}}/\ell_{\mathrm{ad}}$. The rescalings used are described in the main text.

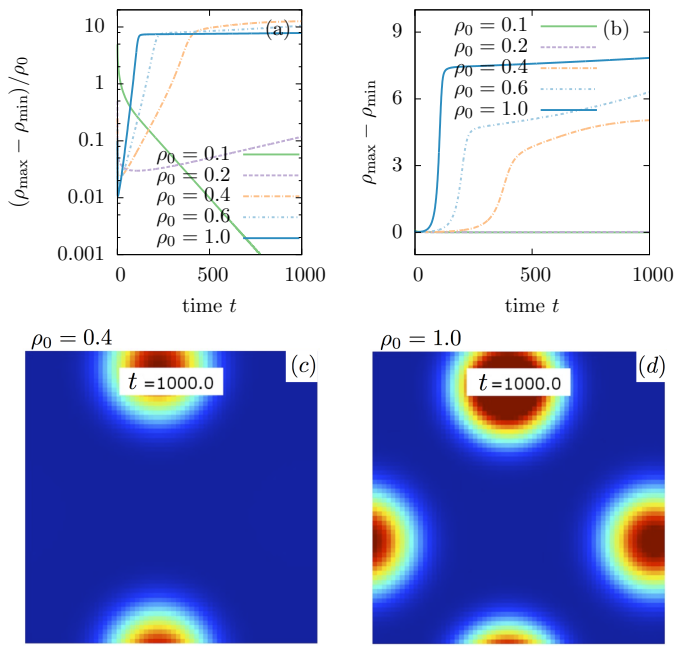

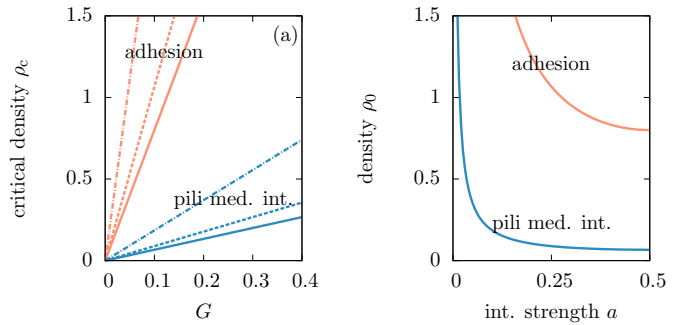

FIG. S2. (color online) (a,b) Maximal density minus minimal density, $\rho_{\max} - \rho_{\min}$, as a function of time $t$, where $\rho_{\max}(t) = \max_{\mathbf{r}}\rho(\mathbf{r}, t)$ and $\rho_{\min}(t) = \min_{\mathbf{r}}\rho(\mathbf{r}, t)$. In (a) we divide through the mean density, $\rho_0 = L^{-2}\int d\mathbf{r}\rho(\mathbf{r}, t)$, where $L$ is the system size (here we used $L = 20$). Note that for all numerical runs, $\rho$ is conserved. For the numerical results (a-d) we used $a = 0.5$, $G = 0.1$ and consider pili-mediated interactions ($\tilde{c}_2 = 3/4$), implying a critical density of $\rho_c = 2/15 = 0.13\overline{3}$ (see main text for the analytic expression of the critical density). (a) Consistently, realizations with $\rho > \rho_c$ exhibit an instability, i.e. $\rho_{\max} - \rho_{\min}$ grows as a function of time, whereas for $\rho_0 < \rho_c$, weak initial spatial perturbations around $\rho_0$ decay exponentially. The growth roughly follows an exponential with a speed strongly dependent on the difference of $\rho_0$ to the critical density: The larger this difference, the faster the initial growth speed. Please note that for densities slightly above the critical threshold (e.g. $\rho_0 = 0.2$), the instability grows too slowly to capture the long time behavior. However, for large enough densities (e.g. $\rho_0 = \{0.4, 0.6, 1.0\}$), $\rho_{\max} - \rho_{\min}$ clearly indicates a saturation. (b) The behavior at very large time-scales is very hard to capture numerically: Interestingly, if the system has developed to a single colony, $\rho_{\max} - \rho_{\min}$ becomes flat (see $\rho_0 = 0.4$ and snapshot (c)). If there are two or more colonies in the system (e.g. $\rho_0 = \{1.0\}$ and snapshot (d)), $\rho_{\max} - \rho_{\min}$ still changes as a function of time, though very weakly. The details of this coarsening mechanism close to saturation will be studied elsewhere.

## S7. Estimates of parameters for *N. gonorrhoeae*

### 1. Estimate for density $\rho_0$

We considered 20 experimental realizations of *N. gonorrhoeae* bacteria forming colony in a substrate. After sedimentation to the plastic surface we calculated from the corresponding binary images the overall area

FIG. S3. (color online) (a,b) Critical density $\rho_c$ as a function of the non-dimensional parameter $G$ and interaction strength $a$ for pili-mediated interactions (blue) and adhesion (red). In (a), three values of interaction strength $a$ are displayed: $(0.5, 0.25, 0.1)$=(solid,dashed,dash-dotted), $\ell_{\mathrm{pi}} \approx 2 \cdot \ell_{\mathrm{ad}}$ and in (b), $G = 0.1$.

fraction covered by bacteria cells, finding $\phi \approx 0.1$. The corresponding dimensionless density is then: $\rho_0 = \phi\ell_{\mathrm{pi}}^2/(\pi R_{\mathrm{cell}}) \approx 0.125$ with $\ell_{\mathrm{pi}} = 1\mu m$ and $R_{\mathrm{cell}} = 0.5\mu m$. Since typically some small three dimensional colonies have already formed during sedimentation process we expect that the determined value represents a slight underestimation, thereby we use $\rho_0 = 0.2$ in the manuscript.

### 2. Estimation of $\mathcal{W}_0/\ell_{\mathrm{pi}}^2$

$\mathcal{W}_0/\ell_{\mathrm{pi}}^2$ is a measure for the rate of cell-cell encounters occurring in an area of $\ell_{\mathrm{pi}}^2$. A direct measurement of this quantity requires the sampling of the cell trajectories on the time-scale of the cell-cell encounters which is intricate because bacteria cells get harmed in case of too frequent light exposure. Therefore, we have to content with a rough estimate. If we assume that each pilus per cell acts independently, we can first estimate the rate of a cell-cell interaction for a single pilus. For intermediate and large cell densities with respect to the intersection scales, it is expected that this rate roughly scales with the number of pili per cell.

The time $\tau$ between two interactions using a single pilus should be roughly given by the time to diffuse the distance to the next-neighbouring cell $\ell_{\mathrm{NN}}$, $\tau \sim \ell_{\mathrm{NN}}^2/(4D)$, where $D = 0.5\mu m^2/s$. However, this distance is reduced by the cell-diameter $2R_{\mathrm{cell}} = 1\mu m$ and two times the typical pili-length $2\ell_{\mathrm{pi}} = 2\mu m$ [9], leading to: $\tau \sim (\ell_{\mathrm{NN}} - 2R_{\mathrm{cell}} - 2\ell_{\mathrm{pi}})^2/(4D) \approx 2s$, where we determined $\ell_{\mathrm{NN}} \approx 5\mu m$ from the binary images directly after sedimentation.

TEM-images have revealed a mean pili number in order of $N \sim 10$ [10]. For intermediate and large cell densities, the overall-interaction rate for for *N. gonorrhoeae* at surface coverage of $\phi \approx 0.1$ is approximately $\mathcal{W}_0/\ell_{\mathrm{pi}}^2 \sim N \cdot (2s)^{-1} \approx 5s^{-1}$. Thereby, $G \approx 0.1$.

## S8. Video Material

Using the procedure mentioned in Section S4 we solved the hydrodynamic equation for various parameters sets.

For a representative parameter set ($G = 0.1$; $a_{ad} = 0.0$; $a_{pi} = 0.5$; $\rho_0 = 0.6$), a video is appended ($T$ denotes the dimensionless time; see main text for corresponding rescaling):

*Colony_formation_via_kinetic_description.mp*4.



\end{document}